\documentclass[a4paper,11pt]{article}
\pdfoutput=1 
\usepackage{jheppub} 

\usepackage{color}
\usepackage{amsmath}
\usepackage{amsfonts}
\usepackage{graphicx}
\usepackage{amssymb}
\usepackage{bbm}
\usepackage{float}
\usepackage{comment}
\usepackage{appendix}
\usepackage[dvipsnames]{xcolor}
\definecolor{refs}{RGB}{245,156,74}
\AtBeginDocument{%
  \let\oldtexttt\texttt
  \renewcommand{\texttt}[1]{\oldtexttt{\detokenize{#1}}}
}

\hypersetup{
  colorlinks = true,
  linkcolor = red,
  urlcolor  = magenta,
  citecolor = blue,
  anchorcolor = blue
}

\newcommand{\dd}{{\rm d}}

\newcommand{\Lag}{\mathcal{L}}
\newcommand{\ham}{\mathcal{H}}
\newcommand{\mH}{\mathcal{H}}

\newcommand{\mS}{\mathcal{S}}
\newcommand{\mQ}{\mathcal{Q}}

\newcommand{\be}{\begin{equation}}
\newcommand{\ee}{\end{equation}}
\newcommand{\bea}{\begin{eqnarray}}
\newcommand{\eea}{\end{eqnarray}}

\renewcommand{\bf}[1]{{\textbf{#1}}}
\renewcommand{\tt}[1]{\textit{#1}}

\newcommand{\mpl}{M_{\rm Pl}}

\newcommand{\xit}{\bar{\xi}}

\begin{document}

\title{Moving superfluids in the rotating universe}

\author[a,b]{Jose Beltr\'an Jim\'enez,}
\author[c]{Federico Piazza,}
\author[a]{Javier Vecino}

\affiliation[a]{Departamento de F\'isica Fundamental and IUFFyM, Universidad de Salamanca, E-37008 Salamanca, Spain.}

\affiliation[b]{Institute of Theoretical Astrophysics, University of Oslo, N-0315 Oslo, Norway.}

\affiliation[c]{Aix Marseille University, Universit\'e de Toulon, CNRS, CPT, Marseille, France.}

\emailAdd{\text{jose.beltran@usal.es}}
\emailAdd{\text{piazza@cpt.univ-mrs.fr}}
\emailAdd{\text{javier.vecino@usal.es}}

\abstract{We study homogeneous cosmological models featuring  
shift-symmetric scalar fields (or, superfluids) in relative motion. In the presence of anisotropy this universe generally features \emph{rotation}, in the sense that the principal axes of anisotropic expansion rotate with respect to the cosmic comoving frame. We focus in particular on the minimal case of two superfluids in 2+1 dimensions. The momentum constraint enforces their spatial gradients to be collinear and the dynamics tends to align such a direction with that of maximal expansion at late times. As opposed to the recently studied case of \emph{solids}, rotation plays a more important role in the present two-superfluids model. The associated energy density does not dilute away but scales as that of anisotropy and affects the total equation of state. We find that purely non-rotating solutions correspond to an unstable surface in phase space in the direction of non-vanishing rotation. This suggests that rotation is a crucial feature of these scenarios that cannot be neglected.}
\date{\today}


\date{\today}
\maketitle

\section{Introduction}

Cosmological models are generically based on the realisation of homogeneity and isotropy as the symmetries of the universe.
These symmetries can be implemented in a straightforward way---at the level of the metric \emph{and} of the fields of the theory---or in a more subtle way, at the level of the metric only, in virtue of the fact that the energy momentum tensor $T^\mu{}_\nu$ feeding the metric is generally more symmetric than the field configurations it is evaluated on. This simple observation has far reaching implications. For example, in the study of black hole solutions, static black holes can be dressed with a time-dependent Galileon field~\cite{Babichev:2013cya}, thereby circumventing standard no-hair theorems. The trick is that time-translational invariance is realised on the Galileon field as a combination of time translations and an internal shift-symmetry. Systems of particular interest are those where Lorentz boosts are spontaneously broken but homogeneity and isotropy are approximately recovered on large scales. Then, a formal way to implement this idea is to consider effective field theories for matter fields with some internal symmetry, and a configuration in which internal and spacetime symmetries are broken down to a diagonal subgroup that preserves homogeneity and isotropy~\cite{Nicolis:2015sra}. Different symmetry breaking patterns identify different universality classes for models of inflation and dark energy---but also of condensed matter physics, which, effectively, corresponds to the limit in which gravity is decoupled. 

Giving up isotropy (while maintaining homogeneity) clearly opens up even more possibilities, with an interesting potential twist: rotation. Anisotropic homogeneous solutions are characterized by different expansion rates along some principal axes. In the presence of anisotropic stresses, these axes can rotate with respect to the comoving cosmological frame~\cite{Nicolis:2022gzh}. In~\cite{Nicolis:2022gzh}  this phenomenon has been studied for 
 \emph{solid matter}~\cite{Soper:1976bb,Dubovsky:2005xd,Endlich:2010hf}. 
 The aim of the present work is to extend these findings to the case of moving superfluids described by shift symmetric scalar fields. The fluids are ``moving" in the sense that the scalar fields have non-vanishing spatial gradients in the cosmological reference frame that give rise to a non-vanishing momentum density $T^0{}_i$ for individual components.  However, spatial homogeneity for the metric field can still be maintained, thanks to the above mentioned enhanced symmetry of the energy momentum tensor.  

Cosmological scenarios with moving components have been considered in e.g. \cite{Maroto:2005kc,BeltranJimenez:2007rsj,BeltranJimenez:2008rei,Garcia-Garcia:2016dcw,Cembranos:2019plq,Cembranos:2019jlp} and the particular case of shift-symmetric scalar fields has been analysed in \cite{Orjuela-Quintana:2024qfn}. In all these cases, the motion occurs along one preferred direction and the metric was assumed to be described by an axisymmetric Bianchi I metric, i.e., only expansion and shear were considered, but rotation was left out. When rotation is brought in, the cosmological dynamics can get substantially more complicated. In this work we will take a first step in this direction. We think that some interesting novel aspects of the multiple fluids rotating model are already encoded in the two-scalar fields example in $2+1$ dimensions to which this paper is devoted. This system bares some important differences with respect to solids and features an interesting unusual dynamics.

Firstly, unlike the solids' case, the scalar fields configuration is no longer static so both time- and space-gradients play a role in the evolution of the universe. 
Secondly, the symmetry breaking pattern is different than that of the solid. In $d$ spatial dimensions, the gravitational sector described by the Einstein-Hilbert action evaluated on homogeneous metrics enjoys an $SL(d,\mathbb{R})$ (global) symmetry, a mini-superspace residual of diffeomorphisms invariance. In the presence of homogeneous scalar fields, this symmetry is not broken. However, a solid breaks it down to its $SO(d)$ subgroup. With the two moving superfluids considered in this paper the original $SL(d,\mathbb{R})$ symmetry of the gravitational sector is broken down to the little group of the unit vector aligned with the spatial gradients of the superfluids. This turns out to be the inhomogeneous special group in $d-1$ dimensions. 
 
Finally, the dynamics of the present system looks richer than that of the rotating solid. In the solid case, the energy density associated with rotation rapidly redshifts away with the expansion. The present system is characterized by two phases. At early times, close to the singularity, the spatial gradients of the fields are negligible, the energy in the rotation becomes subdominant while approaching the singularity and the angle of rotation tends to a constant. As for the shear, it behaves as a kinetically dominated scalar, as in the absence of rotation.  At late times the spatial gradients become important. The rotation angle tends to zero but the energy density associated with rotation scales at the same rate as that associated with anisotropy, and the effective equation of state in 2 dimensions is $w = 1/3$ (this is \emph{not} the same as radiation, which in 2+1 scales with $w=1/2$). Moreover, we find that the small rotation case and that with \emph{strictly zero} rotation appear to behave in a qualitatively different way. This is due to the fact that the non-rotating solutions live on a saddle surface in phase space that is unstable in the directions where rotation is turned on. This behaviour indicates that rotation seems a relevant feature that should not be ignored in the presence of two fluids in relative motion.

 The paper is organized as follows. We will introduce our general set-up in Section \ref{Sec:Generalsetup} where we will write down the mini-superspace action in the presence of an arbitrary number of shift-symmetric scalar fields and in $d$ dimensions. We will discuss the symmetries broken by the matter sector and the important role played by the shift that imposes the momentum constraint. We will then quickly move to discussing the specific model with two canonical scalar fields in 2+1 dimensions in Section \ref{Sec:2scalars}. In that Section we will also discuss the unbroken symmetries and the conserved charges. In Sec.\ref{Sec:DynamicalAnalysis} we will perform the analysis of the dynamical evolution of the system in the presence of both shear and rotation. We will finally discuss our main findings in \ref{Sec:Discussion}.

\section{General setup}
\label{Sec:Generalsetup}

Let us consider homogeneous cosmologies parameterised in ADM variables as
\be
\dd s^2=-\left[N^2(t)-\vec{N}^2(t)\right]\dd t^2+2N_i(t)\dd x^i\dd t+h_{ij}(t)\dd x^i\dd x^j\,,
\label{eq:metric}
\ee
with $N(t)$ the lapse function, $N_i(t)$ the shift and $h_{ij}(t)$ the homogenous metric of the spatial hypersurfaces. As discussed in \cite{Nicolis:2022gzh}, this metric has a global GL$(d,\mathbb{R})$ invariance in $d+1$ dimensions. At the level of the action, the volume element breaks dilations invariance, so only the subgroup SL$(d,\mathbb{R})$ of volume preserving transformations is a symmetry of the gravitational sector. These symmetries are manifest from the gravitational Einstein-Hilbert action evaluated on the homogeneous metrics:
\be
\mS_{\text{GR}}=\frac{\mpl^2}{8}\int\dd t \dfrac{\sqrt{h}}{4N}\left( h^{il}h^{jm}-h^{ij}h^{lm}\right)\dot h_{ij} \dot{h}_{lm}\,.
\ee
Because of diffeomorphism invariance, this sector does not depend on the shift $N_i$, which will however appear in the matter sector to enforce the momentum constraint. The action of this gravitational sector is clearly invariant under a global transformation $h_{ij}\to (\mathcal{A}^{-1})_i{}^m(\mathcal{A}^{-1})_j{}^nh_{mn}$, provided $\det\hat{\mathcal{A}}=1$, and this symmetry will have the associated conserved charges discussed in detail in \cite{Nicolis:2022gzh} (see also below). Such symmetries are nothing but a residual diffeomorphism $x^i\to\mathcal{A}_j{}^ix^j$ corresponding to a global SL$(d,\mathbb{R})$ transformation and, as such, it will act non-trivially on the shift as $N_i\to(\mathcal{A}^{-1})_i{}^jN_j$ as well. This observation is irrelevant for the gravitational sector, which does not contain the shift so, in fact, we could even make it to transform trivially, but it will be important when considering matter sectors containing the shift.

The gravitational sector is identical to the one considered in \cite{Nicolis:2022gzh}. As for the matter sector, here we  assume it is made by $n$ independent shift-symmetric scalar fields $\phi_a$, described by the following Lagrangian density: 
\begin{equation}
{\cal L} = N \sqrt{h} \, \sum_{a = 1}^n P_a(X_a)\,,
\end{equation}
with $X_a=-\frac12 (\partial\phi_a)^2$. As advertised, we are concerned with field configurations with constant spatial gradients, 
\bea
\langle\phi_a\rangle=\phi_a(t)+\vec{\lambda}_a\cdot\vec{x}
\label{eq:backconfig}
\eea
describing $n$ superfluids in relative motion. This configuration is clearly inhomogeneous, but the shift symmetries restore a diagonal translational symmetry realised as $\vec{x}\to\vec{x}+\vec{x}_0$ in combination with $\phi_a\to\phi_a-\vec{\lambda}_a\cdot\vec{x}_0$. A pertinent question at this point is what subgroup of SL$(d,\mathbb{R})$ remains a symmetry of the system. If we perform a transformation $x^i\to\tilde{x}^i=\mathcal{A}_j{}^ix^j$ with $\hat{\mathcal{A}}\in\text{SL}(d,\mathbb{R})$, the configurations \eqref{eq:backconfig} change as
\be
\langle\phi_a\rangle\to\phi_a(t)+(\hat{\mathcal{A}}\cdot\vec{\lambda}_a)\cdot\vec{x},
\ee
so we need $\hat{\mathcal{A}}\cdot\vec{\lambda}_a=\vec{\lambda}_a$ for all $a$ for the system to remain invariant. This means that the space spanned by all the spatial gradients $\vec{\lambda}_a$ need to be an invariant space of $\hat{\mathcal{A}}$, i.e., the remaining symmetries will be the little group of the space generated by the relative motion of the superfluids. Thus, depending on the number $n$ of superfluids and on the configuration of their spatial gradients ${\vec \lambda}_a$, the symmetry group SL$(d,\mathbb{R}$) of the gravitational part of the action can be broken down to a subgroup or to nothing. If the ${\vec \lambda}_a$ only span a subspace of $\mathbb{R}^d$, there is in general a subgroup of SL$(d,\mathbb{R}$) which is still a symmetry of the entire system---the little group associated to this subspace.  If instead the ${\vec \lambda}_a$ are enough to span the entire $\mathbb{R}^d$, then SL$(d,\mathbb{R}$) is entirely broken by the matter sector and no symmetry is left. We defer a more detailed characterisation of the residual symmetry to Appendix \ref{App1} and will now proceed to discussing the momentum constraint enforced by the shift.

In order to characterize our matter sector, we shall first compute the essential building blocks $X_a$ for the configuration \eqref{eq:backconfig} and evaluate them on our homogeneous metric \eqref{eq:metric} to find the following expression:
\bea
    X_a &=& \frac{1}{2N^2} \left( \dot \phi_a -N_i \, \lambda^i_a\right)^2 - \frac{1}{2}h_{ij}\lambda^i_a \lambda^j_a\,,
\eea
with no summation over $a$ in the last term. This expression clearly shows how the spatial gradients couple to both the shift and the spatial metric. The coupling to the shift provides a non-trivial momentum constraint, as we will see below, while the coupling to $h_{ij}$ drives the effects on shear and rotation evolution.

The dynamics of the scalar fields can be described by exploiting the shift symmetries of the scalar fields and their $n$ associated conserved current densities $J_a^\mu=\frac{\partial\Lag}{\partial\partial_\mu\phi_a}$ that satisfy $\partial_\mu J^\mu_a=0$. On the configuration~\eqref{eq:backconfig} it is easy to verify that $J_a^i$ are constant in space, since $X_a$ are homogeneous. It then follows that $\partial_\mu J^\mu_a=\partial_0J^0_a$ so current conservation simply implies a conservation in time of the $n$ charges
\bea
J^0_a\equiv\frac{\partial\Lag}{\partial \dot{\phi_a}}=\frac{\sqrt{h}}{N} P'_a(X_a)  \Big(\dot \phi_a -N_k  \lambda_a^k\Big).
\eea
It is now time to discuss the momentum constraint. The shift only appears in the matter sector and one can see that the following relation holds:
\be
\frac{\partial\Lag}{\partial N_i}=-\frac{\partial\Lag}{\partial\dot{\phi}_a}\lambda^i_a\,,
\label{eq:Nlambdader}
\ee
so the momentum constraint $\frac{\partial {\cal L}}{\partial N^i} = 0$ gives
\begin{equation} \label{momconstr}
\sum_{a = 1}^n \, J^0_a \, \vec{\lambda}_a =0 \,.
\end{equation}
This constraint means that the charge-weighted sum of the gradients of the scalar fields should be zero.\footnote{The fact that we can express the constraint \eqref{momconstr} in terms of the conserved charges can be traced to the existence of a residual symmetry \cite{Orjuela-Quintana:2024qfn} corresponding to $x^i\to x^i-t\epsilon^i$ for a constant $\epsilon^i$. This is nothing but a Galilean boost that we would expect to be a symmetry because only the relative motion among the different components is physical. Under this transformation we have $\delta N_i=\epsilon_i$ and $\delta\phi_a=t\vec{\epsilon}\cdot\vec{\lambda}_a$, while $\delta N=\delta h_{ij}=0$. Given these transformation rules, we see that the invariant building blocks for the mini-superspace Lagrangian is $\dot{\phi}_a-N_i\lambda^i_a$, which explains the relation \eqref{eq:Nlambdader} and, then, the fact that the momentum constraint \eqref{momconstr} can be expressed in terms of the conserved charges. The same relation was shown to hold for a more general shift-symmetric Horndeski Lagrangian in \cite{Orjuela-Quintana:2024qfn}.} An interesting feature of the momentum constraint is that, as the $J^0_a$ are constant, the constraint is time-independent and, in particular, this means that it simply establishes a relation between the charges and the relative motions or, equivalently, for the initial conditions of the system. One simple corollary of~\eqref{momconstr} is that we need at least two fields, in which case $\vec{\lambda}_1$ and $\vec{\lambda}_2$ are aligned and with opposite directions. 

Once we have obtained the momentum constraint associated to the shift, we are free to set the latter to zero. The choice $N_i=0$ is sometimes referred to as the cosmic center of mass frame \cite{Maroto:2005kc,BeltranJimenez:2007rsj}. Notice that this remarks a difference with respect to the case of the rotating solid. There, in the minisuperspace limit, the momentum constraint simply implies $N_i =0$ so it is legitimate to set the shift to zero in the mini-superspace action from the onset. Here, we \emph{can} only set $N_i=0$ after the momentum constraint is enforced. We will stick to this gauge choice from now on.

After presenting our general set-up, we will proceed with the simplest scenario of two canonical scalar fields in 2+1 dimensions. Although this is arguably a simplistic model, it will allow us to illustrate the novel features of these scenarios without unnecessary mathematical complications that may obscure the mechanisms at work. We will see however that already this simple model presents a rich dynamics. 

\section{Two canonical scalar fields in 2+1 dimensions}
\label{Sec:2scalars}

The case of two scalar fields is particularly simple because the momentum constraint~\eqref{momconstr} prescribes that the gradients be aligned. We can then chose to align them along e.g. the $\hat x$ axis at some initial time and they will remain aligned with that axis throughout the entire evolution. From now on, we will assume $\vec{\lambda}_1=\lambda_1 \, \hat{x}$ and $\vec{\lambda}_2=\lambda_2 \,  \hat{x}$, with the momentum constraint implying $\lambda_2 J^0_2=-\lambda_1 J^0_1$. In \cite{Orjuela-Quintana:2024qfn}, this configuration with two fields was used to describe anisotropic cosmologies with shear. Now we extend that study to include rotation.

Before descending to the two dimensional case, let us pause briefly to notice that the remaining symmetries of this configuration in $d$ spatial dimensions correspond to the little group of the preferred direction provided by the the scalar fields spatial gradients, i.e., the superfluids relative motion. In the chosen frame where the motion is aligned with the first axis $\hat{x}$, the little group is given by the elements $\mathcal{A}\in \text{SL}(d,\mathbb{R})$ of the form
\be
\mathcal{A}(\hat{\mathcal{R}},\vec{v})=\begin{pmatrix}
    1&\vec{v}\\
    0&\hat{\mathcal{R}}
\end{pmatrix}
\ee
with $\vec{v}$ an arbitrary $(d-1)$-dimensional vector and $\hat{\mathcal{R}}\in\text{SL}(d-1,\mathbb{R})$. These elements are nothing but a realisation of the special affine group in $d-1$ dimensions so the little group of the residual symmetry is $\text{ISL}(d-1,\mathbb{R})\cong \text{SL}(d-1,\mathbb{R})\ltimes\mathbb{R}^{d-1} $. Thus, the spatial gradients of the scalar fields associated to their relative motion induce a symmetry breaking pattern SL$(d,\mathbb{R})\to$ISL$(d-1,\mathbb{R})$. 
The little group becomes minimal in $d=2$ dimensions where it is one dimensional. We will come back to this in Sec. \ref{Sec:Moreonosymmetry}. Now, after this little digression, let us return to our main focus.

For concreteness, we also specialise to the minimal number of (interesting) spatial dimensions and consider a $2+1$ dimensional spacetime. This will also permit us to straightforwardly compare with~\cite{Nicolis:2022gzh}, where also the 2+1 dimensional case is analysed, to pinpoint the novel features generated by the spatial gradients. The spatial metric in this case is conveniently parameterised as the product of a dilation, a rotation and a shear as follows:
\begin{align}
    h_{ij}=a^2
    \begin{pmatrix}
        \cos\frac{\theta}{2}&\sin\frac{\theta}{2}\\
        -\sin\frac{\theta}{2}&\cos\frac{\theta}{2}
    \end{pmatrix}
    \begin{pmatrix}
        e^\xi&0\\
        0&e^{-\xi}
    \end{pmatrix}
    \begin{pmatrix}
        \cos\frac{\theta}{2}&-\sin\frac{\theta}{2}\\
        \sin\frac{\theta}{2}&\cos\frac{\theta}{2}
    \end{pmatrix},
    \label{eq:2Dmetric}
\end{align}
where $a(t)$ is the isotropic expansion, $\xi(t)$ is the anisotropic expansion (shear) and $\theta(t)$ describes rotation. In terms of this parameterisation, the gravitational mini-superspace action reads
\be
\mS_{\text{GR}}=\frac{\mpl^2}{2}\int\frac{\dd t}{N}\left[-\dot{a}^2+\frac{a^2}{4}\left(\dot{\xi}^2+\sinh^2\xi\dot{\theta}^2\right)\right].
\ee
This sector will then have the symmetries already discussed at length in \cite{Nicolis:2022gzh} corresponding to a global SL$(2,\mathbb{R})$ with the associated conserved charges. The generators of this symmetry can be chosen as
\be
\ell_2=\begin{pmatrix}
    0&1\\-1&0
\end{pmatrix},\quad
\ell_3=\begin{pmatrix}
    1&0\\0&-1
\end{pmatrix},\quad
\ell_4=\begin{pmatrix}
    0&1\\1&0
\end{pmatrix},
\ee
that generate rotations, $+$ shear and $\times$ shear respectively. The associated conserved charges are given by
\bea
Q_2&=&\frac{a^2}{N}\sinh^2\xi\,\dot{\theta},\nonumber\\
Q_3&=&\frac{a^2}{N}\big(\dot{\xi}\cos\theta-\dot{\theta}\sin\theta\sinh\xi\cosh\xi\big),\nonumber\\
Q_4&=&\frac{a^2}{N}\Big(\dot{\xi}\sin\theta+\dot{\theta}\cos\theta\sinh\xi\cosh\xi\Big).
\label{eq:charges}
\eea
At this point it is worth stressing that the conservation of these charges will be subject to the matter sector. In particular, if the matter sector breaks some (or all) of the SL(2,$\mathbb{R}$) symmetries, the charges will no longer be conserved. This is precisely what occurs in our scenario with two perfect superfluids in motion, so let us see how this comes about.

\subsection{Anisotropy equations and conserved charges}

The total mini-superspace action for the system gravity + fields reads
\bea
\mS&=&\frac{\mpl^2}{2}\int\frac{\dd t}{N}\left[-\dot{a}^2+\frac{a^2}{4}\left(\dot{\xi}^2+\sinh^2\xi\dot{\theta}^2\right)+\frac{a^2}{\mpl^2}\left(\dot{\phi_1}^2+\dot{\phi}_2^2\right)\right]\nonumber\\
&&-\frac{\lambda_1^2+\lambda_2^2}{2}\int\dd tN\Big(\cosh\xi-\cos\theta\sinh\xi\Big).
\label{eq:totalaction}
\eea
where we remind the reader that the superfluids motion has been aligned to the $\hat x$ axis. 
The first line is the sector that remains in the absence of motion and reproduces a straightforward generalisation of the scalar matter case studied in \cite{Nicolis:2022gzh} that enjoys the full $\text{SL}(2,\mathbb{R})$ symmetry. On the other hand, the second line represents the effects of the spatial gradients describing the relative motion. Although not obvious at this point, the combination in the second line is precisely the invariant combination under the little group, as we will see below.

The equations for the angle and the shear can be written, in conformal time $\eta$ wherein $N(\eta)=a(\eta)$, as
\begin{align} 
& \frac{\dd^2 \theta}{\dd\eta^2}+\left(\ham+2\frac{\cosh\xi}{\sinh\xi}\frac{\dd \xi}{\dd\eta}\right)\frac{\dd \theta}{\dd\eta}+q_\lambda^2\frac{\sin\theta}{\sinh\xi}=0\,,\label{eq:theta}\\[2mm]
& \frac{\dd^2 \xi}{\dd\eta^2}+\ham\frac{\dd \xi}{\dd\eta}-\cosh\xi\sinh\xi\left(\frac{\dd \theta}{\dd\eta}\right)^2 +q_\lambda^2\big(\sinh\xi-\cos\theta\cosh\xi\big)=0,
\label{eq:shear}
\end{align}
with $\mH=\dd \ln a/\dd\eta $ the Hubble function in conformal time that is governed by the Friedman equation
\be \label{eq:fried}
\mH^2=\frac{q^2}{2a^2}+\frac14\left[\left(\frac{\dd \xi}{\dd\eta}\right)^2+\sinh^2\xi\left(\frac{\dd \theta}{\dd\eta}\right)^2\right]+\frac{q_\lambda^2}{2}\Big(\cosh\xi-\cos\theta\sinh\xi\Big).
\ee
In the above equations we have defined the shifts' and the gradients' charges, 
\be
q^2\equiv (J^0_1)^2+(J^0_2)^2\,,\qquad q_\lambda^2\equiv \lambda_1^2+\lambda_2^2.
\ee
Notice that $q_\lambda$ can be reabsorbed with a redefinition of the time variable $\eta\to\eta/q_\lambda$, provided $q_\lambda\neq0$. Upon this time redefinition, the only remaining parameter is the ratio $\frac{q^2}{q_\lambda^2}$ which only appears in the Friedmann equation. By further rescaling the scale factor $a\to \tfrac{q_\lambda}{q}a$, we can also get rid of the dependence on $\frac{q^2}{q_\lambda^2}$, since $\mH$ is insensitive to a constant rescaling of the scale factor, and, thus, the equations can be recast in a parameter-free form. This permits a global analysis of the dynamics without without having to specify the values of the parameters.

From \eqref{eq:theta} and \eqref{eq:shear} we see that a non-rotating configuration with $\theta=0$ is a consistent truncation of the system, while the truncation with vanishing shear $\xi=0$ necessarily trivialises the rotation as well. In other words, while it is consistent to have shear without rotation, having rotation without shear is not possible. It may be worth emphasising that this does not mean that non-rotating solutions provide attractor solutions of the full system and, as a matter of fact, we will show that non-rotating solutions are actually unstable against rotation perturbations.

Having the equations of motion, we can easily analyse the fate of the charges \eqref{eq:charges}. As already explained, the motion for the superfluids is responsible for the breaking of the symmetries exhibited by the gravitational sector and, therefore, it will induce a violation of the conservation of the charges. This can be checked explicitly by computing the time derivative of the charges and using Eqs. \eqref{eq:theta} and \eqref{eq:shear} to remove second derivatives of the shear and the angle. After doing that, we find the following equations:
\bea
\frac{1}{a}\frac{\dd Q_2}{\dd \eta}&=&-q_\lambda^2\sin\theta\sinh\xi,\\
\frac{1}{a}\frac{\dd Q_3}{\dd \eta}&=&q_\lambda^2\big(\cosh\xi-\cos\theta\sinh\xi\big),\\
\frac{1}{a}\frac{\dd Q_4}{\dd \eta}&=&-q_\lambda^2\sin\theta\sinh\xi.
\eea
The right-hand sides of the above equations arise, as usual, from the variation of the action under the transformation generated by the corresponding charge. For non-moving components $q_\lambda^2=0$, we recover the conservation of these charges because the transformations are symmetries, as shown in \cite{Nicolis:2022gzh}, but they fail to be conserved in the presence of moving components because the symmetries are broken. However, it is obvious to see that the combination
\be \label{eq:charge}
\mathcal{Q}\equiv Q_4-Q_2=a\left[\sin\theta\frac{\dd\xi}{\dd\eta}+\sinh\xi\Big(\cos\theta\cosh\xi-\sinh\xi\Big)\frac{\dd\theta}{\dd\eta}\right]
\ee
remains conserved $\frac{\dd \mathcal{Q}}{\dd \eta}=0$. This is the aforementioned conserved charge associated to the little group of the direction of motion. The transformation corresponds to a combination of a rotation and a shear along the diagonal direction, that is preserved as a residual symmetry because it leaves $\hat{x}$ invariant.\footnote{This is so in the frame where the motion occurs along the $x$-axis. More generally, for an arbitrarily oriented motion, the conserved charge $\mathcal{Q}$ will be a combination of all three charges $Q_1$, $Q_2$ and $Q_3$ and it will generate a shear parallel to the direction of the motion.} Explicitly, it is straightforward to check that $(\ell_2+\ell_4)\cdot\hat{x}=0$.\footnote{The change of sign between $Q_2$ and $\ell_2$ is just a consequence of our definitions of charges and generators.} Our next task will be to analyse the dynamics of the system, but before proceeding to that, let us pause and delve a bit more into the residual symmetry.

\subsection{More on the residual symmetry}
\label{Sec:Moreonosymmetry}

The elements of the little group for our 2+1 dimensional system with two superfluids can be parameterised as
\be
\mathcal{A}(v)=\begin{pmatrix}
    1&v\\0&1
\end{pmatrix},
\ee
with $v$ the parameter of the transformation. The inverse is simply $\mathcal{A}^{-1}(v)=\mathcal{A}(-v)$. These transformations correspond to the parabolic elements\footnote{The elements of SL$(2,\mathbb{R})$ can be classified in terms of the absolute value of their trace into elliptic, parabolic or hyperbolic if such absolute value is smaller, equal or bigger than 2 respectively.} of SL$(2,\mathbb{R})$ and produce a shear mapping of $\mathbb{R}^2$. In order to obtain the explicit transformation for the shear and angle variables we start from our parameterisation \eqref{eq:2Dmetric} and apply the transformation $h_{ij}\to\tilde{h}_{ij}=\Big[\mathcal{A}^T(-v)\,h\,\mathcal{A}(-v)\Big]_{ij}$. Let us first note that we can construct the invariant quantity
\be
\omega^2\equiv\cosh\xi-\cos\theta\sinh\xi,
\ee
that is positive definite.  The invariance of this quantity follows from the invariant sub-space of $\mathcal{A}(v)$, the direction of the motion, that leaves $h_{11}$ invariant. The combination $\omega^2$ is precisely the one that appears in the sector of the action \eqref{eq:totalaction} associated to the motion. We thus see that the action respects the symmetry in a manifest way. 

By comparing the transformation of the trace and the off-diagonal components we find
\begin{eqnarray}
    \cosh\tilde{\xi}&=&\cosh\xi+v\sin\theta\sinh\xi+\frac{v^2}{2}\omega^2,\\
    \sin\tilde{\theta}\sinh\tilde{\xi}&=&\sin\theta\sinh\xi+v\,\omega^2.
\end{eqnarray}
Furthermore, from the transformation of the difference of the diagonal components we obtain
\be
\cos\tilde{\theta}\sinh\tilde{\xi}=(\cos\theta +v \sin\theta)\sinh\xi+\frac{v^2}{2}\omega^2,
\ee
so we can see explicitly that $\omega^2$ in indeed invariant. Assuming that $\tilde{\xi}\neq0$ we can find the following transformation law for the angle:
\be
\tan\tilde{\theta}=\frac{\sin\theta\sinh\xi+v\omega^2}{(\cos\theta+v\sin\theta)\sinh\xi+\frac{v^2\omega^2}{2}}
\ee 
We can see that if we have some values for the shear and the angle $(\theta_0,\xi_0)$ at some time $t=t_0$, we can always do a symmetry transformation to bring the angle to zero by choosing the transformation parameter as
\be
v=-\frac{\sin\theta_0\sinh\xi_0}{\omega_0^2}.
\label{eq:vtheta0}
\ee
This result can be obtained more directly by using a metric parameterised as
\be
 h_{ij}=\begin{pmatrix}
    h_{11}&h_{12}\\h_{12}&h_{22}
\end{pmatrix},
\ee
and applying the transformation
\be
h_{ij}\to\tilde{h}_{ij}=\begin{pmatrix}
    h_{11}&h_{12}-vh_{11}\\h_{12}-vh_{11}&\ h_{22}-2vh_{12}-v^2h_{11}
\end{pmatrix}.
\ee
If we require the transformed metric to be diagonal, that amounts to setting $\theta=0$, we obtain the condition $v=h_{12}/h_{11}$ which, in terms of the shear and the angle, exactly reproduces \eqref{eq:vtheta0}. The component $h_{11}$ is the invariant quantity $\omega^2$ and it is now clear that it expresses the invariance of $h_{ij}\hat{\lambda}^i\hat{\lambda}^j$ where $\hat{\lambda}$ is the unit vector in the direction of the motion.

The fact that we can set the angle to zero at some arbitrary time means that, without loss of generality, we can set initial conditions with vanishing initial angle and, then, the space of solutions can be entirely classified by the initial values of $\xi$, $\dot \xi$ and $\dot \theta$. If we look at the conserved charge \eqref{eq:charge}, that can be used to classify the solutions, we can evaluate it in the representative with $\theta_0=0$ and express it as
\be
\mQ=\frac{a_0}{2}\left(1-e^{-2\xi_0}\right)\left(\frac{\dd\theta}{\dd\eta}\right)_0,
\ee
which shows that the solutions with vanishing charge correspond to either $\xi_0=0$ or $\left(\frac{\dd\theta}{\dd\eta}\right)_0=0$. The former leads to a trivialisation of the equation for $\theta$ so the only viable option for us is having a vanishing initial angular velocity. These are in turn the solutions without rotation that represent a consistent truncation, as discussed above. Thus, solutions with vanishing charge can be mapped into non-rotating solutions and genuinely rotating solutions necessarily have non-trivial charge.

\section{Dynamical analysis}
\label{Sec:DynamicalAnalysis}

For dynamical analysis and numerical integration, it is convenient to use the time variable $x = \log a$ (e-folding number), which grows for an expanding universe and decreases in a contracting phase. In terms of this variable, the Friedmann constraint \eqref{eq:fried} can be expressed as the following cosmic sum rule:
\be
1=\frac{q^2}{2(a\mH)^2}+\frac14\left[\left(\frac{\dd \xi}{\dd x}\right)^2+\sinh^2\xi\left(\frac{\dd \theta}{\dd x}\right)^2\right]+\frac{q_\lambda^2}{2\mH^2}\Big(\cosh\xi-\cos\theta\sinh\xi\Big).
\ee
From this expression, it is apparent that the region of physical solutions for the shear and rotation sector $(\xi,\theta)$ is bound to 
\be
\frac14\left[\left(\frac{\dd \xi}{\dd x}\right)^2+\sinh^2\xi\left(\frac{\dd \theta}{\dd x}\right)^2\right]\leq 1,
\label{eq:physicalregion}
\ee
where the inequality only saturates if the scalar sector trivialises. Since we do not consider that case, our physical region corresponds to the strict inequality. Furthermore, we will see that the boundary of this region is a separatrix.

Regarding the shear and angle variables, eqs.~\eqref{eq:theta}-\eqref{eq:fried} can be written as
\begin{align} \label{4.1}
     \theta''+\left(1 +\frac{1}{\mH^2} \frac{\dd\mH}{\dd\eta}\right) \theta' + 2 \frac{\cosh \xi}{\sinh \xi} \theta' \xi' + \frac{q^2_\lambda}{\mH^2 } \frac{\sin\theta}{\sinh\xi} & = 0\,,\\
      \xi''+\left(1 + \frac{1}{\mH^2}\frac{\dd\mH}{ \dd\eta}\right)\xi' -  \cosh \xi \sinh \xi \, {\theta'}^2 + \frac{q^\lambda_2}{\mH^2} \big(\sinh\xi - \cosh\xi \cos\theta\big)  & = 0\,, \label{4.2}
\end{align}
where a prime denotes derivative with respect to $x$ and we can express
\begin{equation}
    \mH^2 = 2\, \frac{q^2 e^{-2 x} + q_\lambda^2 \big(\cosh\xi - \sinh \xi \cos\theta\big)}{4 - \xi'^2 - \sinh^2\xi  \;\theta'^2},
\end{equation}
\begin{equation}
    1 + \frac{1}{\mH^2}\frac{\dd \mH}{\dd\eta} = \frac{q_\lambda^2}{2\mH^2} \big(\cosh\xi - \sinh \xi \cos\theta\big).
\end{equation}
 The expression for the Hubble function shows that the boundary of \eqref{eq:physicalregion} is indeed a singular surface so that it represents a separatrix, as advertised. The rest of this Section is devoted to understanding the solutions of the above system. In order to build some intuition, we will start by analysing non-rotating solutions with $\theta(t)=0$ which, as discussed, correspond to a consistent truncation. We will later see that this truncation, although consistent, is not stable when rotation perturbations are included.

\subsection{Non-rotating solutions} \label{sec_norotation}

For the non-rotating case, it is convenient to introduce the shifted shear $\xit=\xi-2x$, in terms of which the shear equation for $\theta(t)=0$ reduces to
\be
\xit''=q_\lambda^2\frac{1+\tfrac14\xit'}{q_\lambda^2+q^2e^{\xit}}\xit'^{\, 2}.
\label{eq:nonrotatingshear}
\ee
From this equation we easily recover that, in the absence of spatial gradients for the scalars, $q_\lambda=0$ (i.e., non-moving superfluids), the shear evolves with constant value $\xit'$ determined by initial conditions. For $q_\lambda\neq0$ we can absorb the constants $q$ and $q_\lambda$ into a shift $x\to x-\frac12\ln\frac{q^2}{q_\lambda^2}$ so, without loss of generality, we can set $q^2=q_\lambda^2=1$. We will use these values for our numerical integrations and general arguments, keeping in mind that changing the values of $q^2$ and $q_\lambda^2$ will be equivalent to shifting the origin of $x$. We will however leave arbitrary values in our analytical developments to keep track of the origin of the different features and regimes that we will find.

\begin{figure}[ht]
\centering
\includegraphics[width=0.54\linewidth]{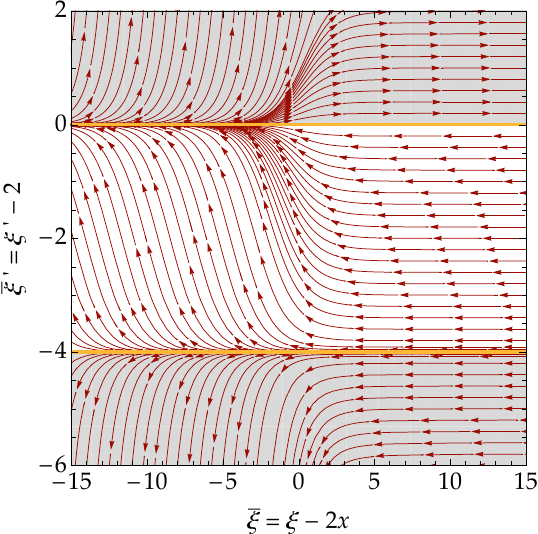}
\includegraphics[width=0.435\linewidth]{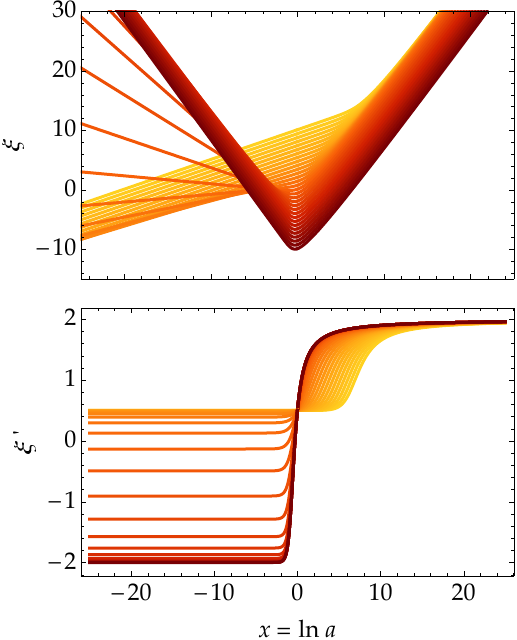}
\caption{\bf{Left panel}: phase map of the non-rotating solutions.
 The arrows indicate the direction of growing $x$. The lines at $\xit'=0$ and $\xit'=-4$ are two separatrices and the white non-shaded area in between represents the physical region with $(a\mH)^2>0$. The trajectories of this region are attracted towards $\xit'=0$ in an expanding phase. \bf{Right panels}: the evolution of the shear for different initial conditions. We have set them at $x=0$ with $\xi'(0)=0.5$ and $\xi(0)$ ranging between $-10$ and $10$.  While the late-time attractor always corresponds to $\xi'=2$ ($\xit'=0$), the early-time attractor depends on the sign of $\xi(0)$. Negative values of $\xi(0)$ correspond to early-time solutions with constant derivative different from the separatix $\xi'=-2$, while negative initial values of the shear have the separatrix as early-time attractor. This is in agreement with our analytical findings.  }
\label{Fig:phasemap}
\end{figure}

Since the independent variable has been completely absorbed into the shear, the equation \eqref{eq:nonrotatingshear} is an autonomous system whose phase map is shown in Fig. \ref{Fig:phasemap}.
We can see that there are two {\it critical} trajectories corresponding to equilibrium solutions with $\xit'=0$ and $\xit'=-4$. These are the boundaries of the (unshaded) physical region for non-rotating solutions given in~\eqref{eq:physicalregion} and which act as separatrices in phase space. The trajectories with $\xit'=0$ describe solutions with constant $\xit$. The stability of these solutions can be studied by expanding $\xit=\xit_0+\delta\xit$, with $\xit_0$ the constant value of the shear, and  the shear equation can be perturbed as
\be
\delta\xit''\simeq q_\lambda^2\frac{{\delta\xit'}^{\, 2}}{q_\lambda^2+q^2e^{\xit_0}}\,.
\label{eq:deltaxiattractor}
\ee
This equation shows that $\delta\xit''$ is always positive, which means that $\delta\xit'$ grows  so that trajectories near $\xit'=0$ flow upwards in phase space. Hence, the trajectories below the separatrix are attracted to it whereas those above the separatrix run away, as we see in Fig. \ref{Fig:phasemap}. If $\xit\gg1$, the exponential factor in \eqref{eq:deltaxiattractor} strongly suppresses the growth of $\delta\xit'$ and the runaway becomes extremely slow. For the separatrix with $\xit'=-4$, the small perturbations are governed by
\be
\delta\xit''\simeq \frac{4q_\lambda^2}{q_\lambda^2+q^2e^{\xit}}\delta\xit'.
\ee
In this case, the sign of $\delta\xit''$ coincides with that of $\delta\xit'$ so the nearby trajectories are always repelled by the separatrix.
Thus, the separatrix $\xit'=0$ is a late-time attractor for the trajectories between the two separatrices, i.e., with $-4<\xit'<0$. This behaviour is clearly seen in Fig. \ref{Fig:phasemap}. In that Figure we can also see that all trajectories with large (positive) $\xit$ tend to be horizontal, i.e., they evolve with constant $\xit'$. This can be easily understood because, in the regime with $\xit\gg1$, the equation for $\xit$ simplifies to
\be
\xit''\simeq \frac{q_\lambda^2}{q^2}e^{-\xit}\left(1+\frac{1}{4}\xit'\right)(\xit')^2
\ee
so the second derivative is exponentially suppressed and hence $\xit'$ remains constant.

Now that we have analysed the evolution of the shear, we can proceed to solving the expansion. For the non-rotating solutions, the Hubble function can be expressed as
\be
(a\mH)^2=-2\frac{q_\lambda^2+q^2e^{-\xit}}{(4+\xit')\xit'}.
\ee
Besides determining the isotropic expansion in terms of the shear, this expression also shows that we need $-4<\xit'<0$ in order to have physical solutions with positive $(a\mH)^2$. This is just the bound \eqref{eq:physicalregion} restricted to non-rotating solutions and we see explicitly here that they indeed are separatrices of the dynamical system. We thus conclude that the non-rotating physical solutions correspond to the trajectories between the two separatrices, and, hence, all physical solutions are eventually attracted towards $\xit'=0$. Near this attractor, the equation for the shear is
\be
\xit''\simeq \frac{q_\lambda^2}{q_\lambda^2+q^2e^{\xit}}\xit'^{\, 2}.
\label{eq:nonrotatingshearseparatrix}
\ee
Since near the separatrix $\xit$ grows very slowly (because $\xit'\simeq0$), we obtain the following approximate solution near the attractor:
\be
\xit'\simeq-\left(1+\frac{q^2}{q_\lambda^2}e^{\xit}\right)x^{-1}.
\ee
At sufficiently large times near the attractor we have $\xit\to-\infty$ so, at leading order, we find $\xit\simeq-\ln x$, which confirms that $\xit$ varies very slowly.\footnote{One needs to be cautious however because $\xi$ is exponentiated in the equations so it can be important to keep track of the correct order in the expansions.} At large times, we then have $\xit'\simeq-\left(1+\frac{q^2}{q_\lambda^2}\frac{1}{x}\right)x^{-1}\simeq-x^{-1}$. Equipped with this solution we can finally obtain the evolution of the Hubble factor as
\be
(a\mH)^2\simeq-\frac{q_\lambda^2+q^2e^{-\xit}}{2\xit'}\simeq \frac{q^2}{2}x^2.
\ee

If we translate this back to the original variable $\xi$, we get that, at early times $(x\ll -1)$, close to the singularity, $\xi'$ becomes a constant of value between $-2$ and $2$. In the absence of gradients there is no preferred spatial direction and switching between the $\hat x$ and $\hat y$ axes amounts to turn $\xi$ into $-\xi$. The limits $\xi'=2$ and $\xi' = -2$ both correspond to $\dot \xi^2 = 4H^2$ which in our units represent a universe totally dominated by anisotropy. Any intermediate value between $-2$ and $2$ instead represents the presence of the time derivatives of the scalars in the Friedmann equation. They scale as anisotropy and therefore share a constant fraction of the entire energy budget. 

At late times $(x\gg 1)$ the spatial gradients of the scalars become important and the symmetry $\xi \rightarrow -\xi$ is broken by the fact that we have conventionally chosen to align such gradients along the $\hat x$ axis. The late time attractor, $\xi'=+2$, means that the direction of motion of the superfluids ends up being that of maximal expansion eventually.  During the evolution it is thus possible that the direction of maximal expansion switches from $\hat y$ to $\hat x$, the switch happening at around the time when gradients become important. 

Let us now turn our attention to the case with rotation.

\subsection{Rotating solutions}

Let us start our analysis of rotating solutions by exploring the space of physical solutions. For that, we can consider the Hubble function that can be expressed as
\be \label{xinew}
(a\mH)^2=2\frac{q_\lambda^2+q^2e^{2x}(\cosh\xi-\sinh\xi\cos\theta)}{4-(\xi'^2+ \sinh^2\xi\,\theta'^2 )}\,.
\ee
Since the numerator is always positive, we need to satisfy the following condition
\be
\xi'^2+\sinh^2\xi\, \theta'^2<4,
\label{eq:physicalregion2}
\ee
which is again the physical region \eqref{eq:physicalregion}. From the expression of the Hubble function we corroborate that the surface $\xi'^2+\sinh^2\xi\, \theta'^2=4$ is a singular surface of the equations so it acts as a separatrix surface. This feature disconnects the physical and non-physical regions so they are not dynamically connected, as one would expect. In particular, we obtain that $\vert\xi'\vert$ can never exceed 2 and this bound is saturated by $\xi=0$ and/or $\theta'=0$. This is precisely the physical region obtained for the non-rotating solutions, which means that including rotation shrinks the physical region for the shear variable $\xi$ with respect to the non-rotating case.

\begin{figure}[ht]
\includegraphics[width=0.49\linewidth]{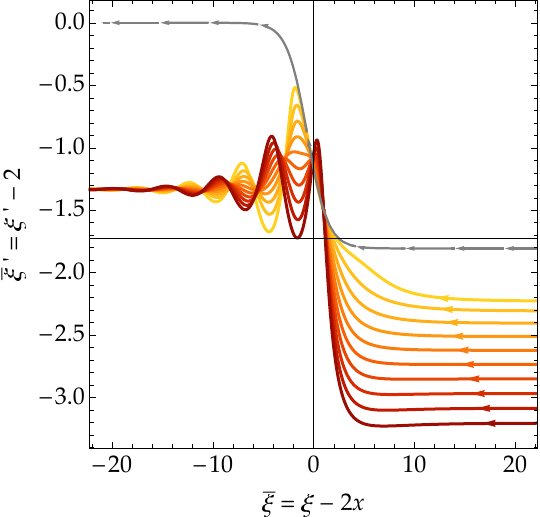}
\includegraphics[width=0.49\linewidth]{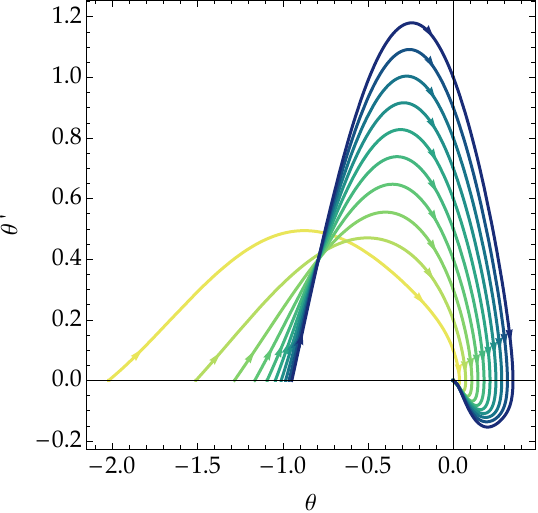}
\includegraphics[width=0.49\linewidth]{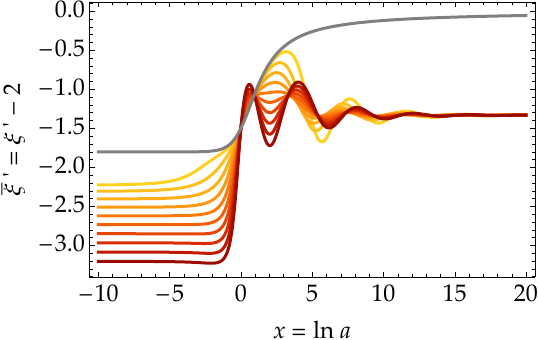}
\includegraphics[width=0.49\linewidth]{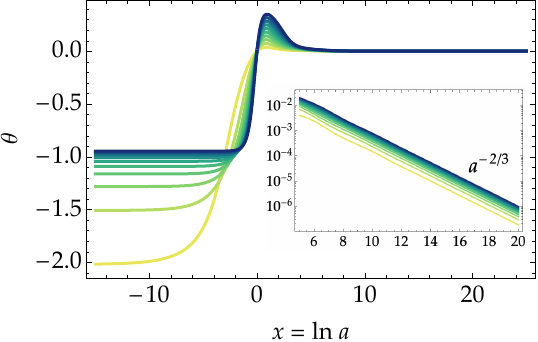}
\caption{In this plot we show the shear and angle evolutions obtained by numerically solving the equations. We have chosen initial conditions $\xi(x=0)=1$ and $\xi'(x=0)=0.5$ for the shear, while for the rotation we have selected initial velocities ranging from $0$ (lighter) to $1$ (darker) and $\theta(x=0)=0$. The initial angle has been fixed to $\theta_0=0$ which is always possible by means of a symmetry generated by $\mathcal{Q}$ as explained in the main text. The upper panels show the phase map trajectories and the lower panels correspond to the time evolution.  We can see how the angle evolves towards zero, but it leaves an imprint in the evolution of the shear, whose asymptotic evolution has $\xit'=-4/3$ ($\xi'=2/3$). This leaves a memory of the rotating phase since the pure non-rotating solution (shown in gray) is attracted towards  $\xit'=0$ ($\xi'=2$).} 
\label{Fig:thetaxiev}
\end{figure}

\begin{figure}[ht]
\centering
\includegraphics[width=0.5\linewidth]{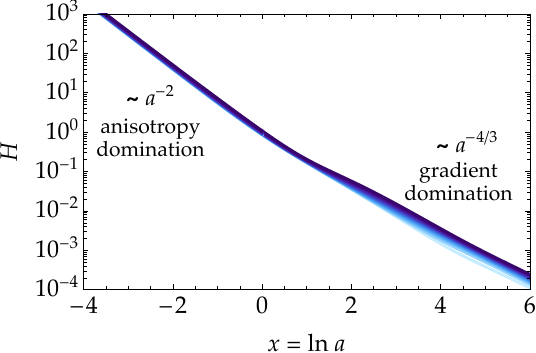}
\includegraphics[width=0.49\linewidth]{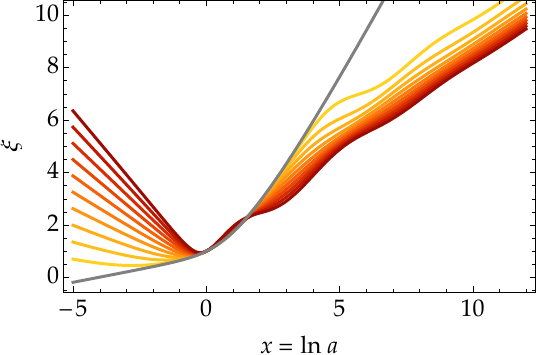}
\caption{Evolution of the Hubble function (left) and the shear (right) for the same set of initial conditions as in Fig. \ref{Fig:thetaxiev}. The gray line corresponds to the non-rotating solution and we see how the presence of rotation crucially changes the asymptotic evolution of the shear.}
\label{Fig:Hubble}
\end{figure}

One way to understand the attractor in the general case is to look again at the charge~\eqref{eq:charge}, whose conservation can be expressed as  
\begin{equation} \label{eq:charge3}
 \frac{\mathcal{Q}}{a\mH}=\sin\theta\, \xi'+\sinh\xi\Big(\cos\theta\cosh\xi-\sinh\xi\Big)\theta'\, .
\end{equation}
 At early times spatial gradients become irrelevant and rotation becomes a subdominant component with the angle $\theta$ stabilizing towards a constant value (see Fig. \ref{Fig:thetaxiev}) approaching the singularity. Not surprisingly, we recover an epoch dominated by anisotropy, with $\xi'\sim$ const. and $H^2 \propto a^{-4}$ (see Fig. \ref{Fig:Hubble}).

When gradients become important at late times, the behaviour of the system is instead relatively unprecedented. 
On the attractor, every term in~\eqref{eq:charge3} scales in the same way. This is achieved\footnote{Here and in what follows we use the small-$\theta$ approximations
\begin{align}
\cos\theta\cosh\xi-\sinh\xi \ &\simeq \ \frac12\left(- \frac{\theta^2}{2} e^{\xi} + 2 e^{-\xi}\right),\\
   \cosh\xi-  \cos\theta\sinh\xi \ &\simeq \ \frac12\left( \frac{\theta^2}{2} e^{\xi} + 2 e^{-\xi}\right).
\end{align}} by the following asymptotic behaviors at $x\rightarrow \infty$:
\begin{align}\label{asy1}
    \theta \ &\simeq \ \theta_\infty e^{-\xi'_\infty x}\, , \\
    \xi \ & \simeq \ \xi_\infty' x\, ,\\
    {\cal H} \ & \simeq \mH_\infty e^{(\xi'_\infty - 1)x}\, , \label{asy3}
\end{align}
for some constants $\theta_\infty$, $\xi_\infty'$ and $\mH_\infty$. At the same time, this is the regime where the gradients dominate. If we insert~\eqref{asy1}-\eqref{asy3} into the expression for the Hubble function \eqref{xinew}, we find the asymptotic constant value of $\xi'$ to be
\begin{equation}
    \xi'_\infty = \frac23\, .
\end{equation}
This asymptotic behaviour is corroborated by the numerical solutions shown in Fig. \ref{Fig:thetaxiev}. The remaining constants depend on the specific initial conditions and can be algebraically related to $\mQ$, although their particular relation is not of interest for us here. Let us finally note that the proper time Hubble function $H=\mH/a$ in this asymptotic regime scales as $H\propto a^{-4/3}$ as also shown in the left panel of Fig. \ref{Fig:Hubble}.

\subsection{Little rotation $\neq$ no rotation}

One remarkable feature appearing from the above analysis is that the presence of rotation, no matter how small, crucially changes the late-time behaviour of the system. In other words, the zero-rotation limit is discontinuous in phase space. The point is that the attractor of the non-rotating solutions obtained in Sec.~\ref{sec_norotation} does not correspond to an attractor in the full phase space, but rather to a saddle surface which is unstable along the additional directions corresponding to rotation. We can see this analytically, by expanding the theta and shear equations eqs.~\eqref{eq:theta} and~\eqref{eq:shear} at small angles $\theta\ll1$ around the non-rotating late-time attractor. As $\xi$ increases along the attractor, eq.~\eqref{eq:theta} is approximated by 
\be
\frac{1}{a}\frac{\dd}{\dd\eta}\left(a\frac{\dd\theta}{\dd\eta}\right)+2\frac{\dd\xi}{\dd\eta}\frac{\dd\theta}{\dd\eta}\simeq0 , 
\ee
which admits solutions with constant angle $\theta$ irrespectively of the evolution of the shear. Of course this is not the behaviour of the angle along the real attractor in the full phase space, eq.~\eqref{asy1}, but just how $\theta$ tends to behave when forced to be close to the non-rotating solution. 

On the other hand, the shear equation~\eqref{eq:shear} at small angles and large $\xi$ implies 
\be
\frac{1}{a}\frac{\dd}{\dd\eta}\left(a\frac{\dd\xi}{\dd\eta}\right)-q_\lambda^2e^{-\xi}\simeq\frac{e^\xi}{4}\left[e^{\xi}\left(\frac{\dd\theta}{\dd\eta}\right)^2-q_\lambda^2\,\theta^2\right].
\ee
As the correction w.r.t. the non-rotating case is quadratic in the angle, we can simply plug in the constant $\theta$ solution obtained before and treat it as a small external parameter. But this is clearly a ``huge" source as compared to the remaining terms of the equation, growing like $e^{\xi}$ (v.s. $e^{-\xi}$ on the LHS) and destabilizing the system away from the non-rotating solution. Curiously, the inclusion of a small rotation takes the system to another attractor where $\theta$ tends to zero and $\xi'$ tends to a constant that is different from that of the strictly non-rotating solution.

\section{Discussion}
\label{Sec:Discussion}

The cosmological scenarios analysed in this work could appear somewhat exotic. However, the presence of two scalar fields is a very straightforward generalization with respect to the single scalar field case and, as a matter of fact, multiple shift-symmetric scalar fields appear naturally as Goldstone bosons arising in phases with broken symmetries.  Once we accept the presence of multiple shift-symmetric scalars, the next natural question is how natural the inhomogeneous configuration that we have considered is. In principle, there is no reason to assume that the homogeneity frame of all the scalar fields should be the same.  As we have discussed, our particular configuration, despite being inhomogeneous, gives rise to a homogeneous energy-momentum tensor and the system can be interpreted as a set of super-fluids in relative motion. Thus, the question can be equivalently rephrased as how natural is for the two superfluids to share a common rest frame, especially in the absence of direct couplings.

Usually, homogeneity is given for granted from the onset, but if some relative velocity is present between the two scalars then, as we have shown, rotation is a relevant dynamical feature at late times that should not be ignored. While the rotation angle tends to zero (which means that the gradient direction aligns with that of maximal expansion), the energy density in the rotation scales like the shear and changes the equation of state of the system which becomes $w = 1/3$ in 2+1 dimensions.\footnote{In 2+1 dimensions radiation has equation of state $1/2$, and the energy density of a component with equation of state $w$ scales as $\rho \propto a^{-2 (w+1) .}$} In this respect, no-rotation is very different than little-rotation.  
One other relevant feature of this model is that the direction of maximal expansion can switch during time evolution. This happens even in the absence of rotation.

Of course it would be interesting to extend this study to 3+1 dimensions. At first glance  doing that in general would imply a substantial amount of complication (see e.g. App. C of Ref.~\cite{Nicolis:2022gzh}). This is not surprising, since the rigid body in 3 spatial dimensions is already relatively complicated, even without considering the additional degrees of freedom of anisotropy and expansion that are present in the cosmological case. Still, one can imagine a straightforward extension of the present model where one basically adds one dimension orthogonal to the plane of the rotation. If no shear is present along the additional dimension, the dynamics of such a 3+1 dimensional model should be a straightforward generalization of the one considered here. These more realistic 3+1 cases clearly deserve further study.

Another remark we should make is that we have considered canonical scalar fields in this work, but our general framework can be applied to completely general shift-symmetric multi-scalar field theories. Although some of our results might be specific of canonical scalar fields, they already put forward the importance of considering rotation in these scenarios. In this respect, we do not see any reason to expect the unstable nature of strictly non-rotating solutions that we have found to extend to more general models. However, our results reveal that rotation is an important ingredient that cannot be neglected.

As a final comment, more closely related to practical applications, the rotating cosmologies supported by shift-symmetric scalar fields could provide a novel class of scenarios with potential relation to dipole anomalies, see e.g. \cite{Perivolaropoulos:2021jda,Secrest:2025wyu}.

\begin{acknowledgments}
JBJ acknowledges support from Projects PID2021-122938NB-I00 and PID2024-158938NB-I00 funded by the Spanish “Ministerio de Ciencia e Innovaci\'on" and FEDER “A way of making Europe”, the Project SA097P24 funded by Junta de Castilla y Le\'on and the research visit grant PRX23/00530. JBJ thanks the Institute of Theoretical Astrophysics at the University of Oslo for their hospitality during the completion of this work.
The work of F.P. received support from the French government under the France 2030 investment plan, as part of the Initiative d'Excellence d'Aix-Marseille Universit\'e - A*MIDEX (AMX-19-IET-012). It was also supported by the ``action th\'ematique" Cosmology-Galaxies (ATCG) of the CNRS/INSU PN Astro and by the {\it Agence Nationale de la Recherche} under the grant ANR-24-CE31-6963-01.
\end{acknowledgments}

\appendix

\section{Residual symmetry in the general case}
\label{App1}
In this Appendix we will give a more detailed characterisation of the little group in the general case. Barring accidental linear dependencies among $\vec{\lambda}_a$ there is at most $n-1$ linearly independent spatial gradients due to the momentum constraint. In this case, the relation between the dimension $d$ and the number of components $n$ determines the residual unbroken subgroup. If $d\leq{n-1}$, $\{\vec{\lambda}_a\}_{a=1}^{n-1}$ will provide a complete basis of $\mathbb{R}^d$ so there is no remaining invariant sub-space under SL$(d,\mathbb{R})$. The situation is more interesting when $n-1<d$ because the set of spatial gradients $\{\vec{\lambda}_a\}_{a=1}^{n-1}$ only span a sub-space of $\mathbb{R}^d$ whose little group determines the unbroken symmetries. We can construct the elements of the little group by taking $\{\vec{\lambda}_a\}_{a=1}^{n-1}$ as the first vectors of a basis in $\mathbb{R}$ so that SL$(d,\mathbb{R})$ breaks down to the subgroup that has $\text{span}\{\vec{\lambda}_a\}_{a=1}^{n-1}$ as invariant sub-space. If we complete this invariant sub-space with a basis in its complementary, the elements of the unbroken sub-group can be parameterised as
 \bea
 \mathcal{A}_{d,n}(\hat{v},\hat{\mathcal{R}})=\begin{pmatrix}
        \mathbbm{1}_{n-1}&\hat{v}\\
       \hat{0}&\hat{\mathcal{R}}
  \end{pmatrix}
\eea
with $\hat{\mathcal{R}}$ a unitary squared $(d-n+1)$ matrix and $\hat{v}$ a $(n-1)\times(d-n+1)$ matrix. Thus, out of the original $(d^2-1)$ symmetries, $d(n-1)$ are broken and $d(d-n+1)-1$ remain. This is the smallest remaining sub-group, but it could be a bigger symmetry if additional linear dependencies between the $\vec{\lambda}_a$'s occur. Let us specify to the interesting cases of $d=2$ and $d=3$.
\begin{itemize}
    \item $d=2$. In two dimensions, the only possibility to have a non-trivial little group is to have $n=2$ components, i.e., two superfluids whose spatial gradients are then aligned in the frame with vanishing shift. The elements of the little group can be parameterised as
\be
\mathcal{A}_{2,2}(v)=\begin{pmatrix}
    1&v\\0&1
\end{pmatrix}
\ee
with $v$ a constant parameter. Intuitively, this transformations simply correspond to the elements of SL$(2,\mathbb{R})$ describing a shear parallel to the direction of the spatial gradients. This case was discussed in Sec. \ref{Sec:Moreonosymmetry}.

\item $d=3$. In three dimensions we have two possibilities, namely: $n=2$ or $n=3$. If $n=2$, the situation is the generalisation of the two-dimensional case to three dimensions since we have again one preferred direction whose little group provides the unbroken symmetries. In the three dimensional case however the little group has three parameters and is described by
\be
\mathcal{A}_{3,2}(\vec{v},\hat{\mathcal{R}})=\begin{pmatrix}
    1&\vec{v}\\
    \vec{0}&\hat{\mathcal{R}}
\end{pmatrix}
\ee
with $\vec{v}=(v_1,v_2)$ a constant vector and $\hat{\mathcal{R}}\in$SL$(2,\mathbb{R})$. Intuitively, this little group can be understood as the special linear transformations in the plane orthogonal to the motion plus shear transformations parallel to the direction of the motion. For $n=3$ instead, the spatial gradients span, in general, a plane of $\mathbb{R}^3$ and the little group can be described by 
\be
\mathcal{A}_{3,3}(\vec{v})=\begin{pmatrix}
    1&0&v_1\\
    0&1&v_2\\
    0&0&1
\end{pmatrix}.
\ee
In this case, the little group is Abelian and simply describes shears parallel to the plane spanned by the motion.
\end{itemize}

\bibliography{BibRotating}

\end{document}